\documentclass[usenatbib]{mn2e} \usepackage{psfig}

\newcommand{\Msun}{\mbox{$\rm M_{\odot}$}}

\newcommand{\LCDM}{\mbox{$\Lambda$CDM}} 
\newcommand{\SCDM}{\mbox{SCDM}} 
\newcommand{\TCDM}{\mbox{$\tau$CDM}}
\newcommand{\generation}{\mbox{Generation}}
 
\newcommand{\Tvir}{\mbox{$T_{\rm vir}$}} 
\newcommand{\Tsf}{\mbox{$T_{0.75}$}}
\newcommand{\rvir}{\mbox{$r_{\rm vir}$}} 
\newcommand{\tvir}{\mbox{$t_{\rm vir}$}} 
\newcommand{\zvir}{\mbox{$z_{\rm vir}$}}

\newcommand{\tcool}{\mbox{$t_{\rm cool}$}}
\newcommand{\tdyn}{\mbox{$t_{\rm dyn}$}}
\newcommand{\dtcoll}{\mbox{$\Delta t_{\rm coll}$}}
\newcommand{\Mtot}{\mbox{$M_{\rm tot}$}}

\newcommand{\htwo}{\mbox{H$_2$}}

\title
  [Primordial star clusters]
  {The mass-function of primordial star clusters}

\author
  [F. Santoro \& P. A. Thomas]
  {Fernando Santoro and Peter A.~Thomas
   \thanks{E-mail: p.a.thomas@sussex.ac.uk}\\
   Astronomy Centre, University of Sussex, Falmer, Brighton, BN1\,9QJ }

\date{\today}

\begin{document}
\journal{Preprint astro-ph/0209152}
 
\maketitle

\begin{abstract}

We use the block model to generate merger trees for the first star
clusters in a $\Lambda$CDM cosmology.  Using a simple collapse model
and cooling criterion, we determine which halos are able to form stars
before being disrupted by mergers.  We contrast the mass functions of
all the resulting star clusters and those of primordial composition,
i.e. star clusters that have not been contaminated by subclusters
inside them.  In confirmation of previous work, two generations of
primordial star clusters are identified: low-temperature clusters
that cool via molecular hydrogen, and high-temperature clusters
that cool via electronic transitions.  The former dominate by number,
but the two populations contain a similar mass with the precise
balance depending upon the details of the model.  We speculate on the
current-day distribution of Population-{\sc iii} stars.

\end{abstract}


\section{Introduction}
\label{sec:intro}

In a Cold Dark Matter (CDM) cosmology, small structures are the first
to collapse and these then cluster together in a hierarchical fashion,
giving rise to the bottom-up picture of galaxy formation.  In this
paper, we use the term `primordial star cluster' for the first objects
that are able to cool and form, zero-metallicity, Population\,{\sc iii}
stars.  They are of interest both in their own right and because they
may be responsible for reionization of the intergalactic medium.

In a primordial gas, whose main elements are hydrogen and helium and
their derivatives, there are two main cooling mechanisms, dependent on
the temperature: for halos with virial temperatures less than
8\,600\,K the cooling is dominated by roto-vibrational excitations of
hydrogen molecules, while those with higher temperatures cool mainly via
electronic transitions.

In a landmark paper entitled ``How small were the first cosmological
objects?'', \citet[][hereafter T97]{TSR97} analytically tracked a
top-hat collapse to the point of virialization, at which point the gas
was cooled at constant density.  They accepted an object as having
cooled if it met the criterion $T(0.75\zvir)\leq0.75\,\Tvir$, where
\Tvir\ is the virial temperature and \zvir\ the virialization
redshift.  They found that the first generation of objects that cooled
in a standard CDM scenario virialized at a redshift of 27 and had a
baryonic mass of about $10^5$\Msun.  In a later paper, \citet{AAN98}
redid the calculation with a different \htwo\ cooling function
and estimated a very similar virialization redshift but a smaller
baryonic mass, $7\times10^3$\Msun.

In a paper somewhat cheekily entitled ``How big were the first
cosmological objects?'', \citep[][hereafter HSTC02]{HST02} extended
the previous work to include halos of higher mass.  They found two
distinct generations of halos: \generation~1 halos dominated by
molecular cooling as in the previous work, and higher-mass,
\generation~2 halos dominated by electronic cooling.  The properties
of these halos in three different versions of the CDM cosmology are
listed in Table~\ref{tab:halo}.

\begin{table}
\caption{Properties of the two generations of 3-$\sigma$ halos to form
in each of the cosmologies (from HSTC02): cosmological model; redshift
at which the halo cools to 75 per cent of the virial temperature;
virial temperature; total mass of the halo; baryonic mass of the
halo.}
\label{tab:halo}
\begin{tabular}{rcccc}
Model&      $z_{0.75}$&   \Tvir/K&   $M_{\rm    tot}/\Msun$&   $M_{\rm
bary}/\Msun$\\
\hline
Generation~1\\
\SCDM& 19.5& 3\,600& $1.6\times10^6$& $1.2\times10^5$\\
\TCDM& 10.8& 4\,500& $5.0\times10^6$& $9.2\times10^5$\\
\LCDM& 21.9& 3\,400& $8.6\times10^5$& $3.3\times10^4$\\
Generation~2\\
\SCDM& 18.5& 10\,800& $9.9\times10^6$& $7.5\times10^5$\\
\TCDM& 10.7& 10\,600& $2.1\times10^7$& $3.9\times10^6$\\
\LCDM& 20.6& 10\,400& $5.7\times10^6$& $2.2\times10^5$
\end{tabular}
\end{table}
Note that the difference in cooling redshift between the two
generations of halos is small (partly because the CDM fluctuation
spectrum is flat on small scales and partly because cooling is more
efficient in \generation~2 halos).  This led HSTC02 to speculate that
both generations of halos may form Population~{\sc iii} stars (whereas
others have considered only the smaller, \generation~1 halos to be
important).

There were two main deficiencies in the model of HSTC02.  Firstly, they
neglected substructure: the referee suggested that all \generation~2
halos will have \generation~1 halos inside them and so they will not
be of primordial composition.  We show below that primordial
\generation~2 halos can exist.  Secondly, they considered only
3-$\sigma$ fluctuations.  In reality there will be a gaussian
distribution of overdensities leading to a wide range of halo masses
virialising at any given redshift.

The present  paper, as a  continuation of the previous work, addresses
these two points by using the Block Model of \citet{CoK88} to generate
a merger history of collapsed halos.  This allows us to follow a wide
dynamic range of halo masses very efficiently.  Within the merger
tree, the properties of halos are calculated using the same chemical
model as in HSTC02.

Our work complements that of other authors who are investigating first
object formation using numerical simulations, e.g. \citet{AAN98},
\citet*{BCL99}, \citet*{ABN00}, \citet*{FuC00}, \citet{NaU01},
\citet*{BCL02} and \citet{NaU02}.  They are able to follow the dynamical
evolution of single star clusters in great detail, whereas we learn
instead about the properties of the cluster population.

We describe our numerical method in Section~\ref{sec:method}.  The
properties of individual halos for a merger tree corresponding to an
overdense region of the Universe are presented in
Section~\ref{sec:properties} and the corresponding mass function is
described in Section~\ref{sec:massfn}.  Finally, Section~\ref{sec:discuss}
explores variations of the basic model and discusses the nature of the
star clusters.

\section{Methodology}
\label{sec:method}

In this paper, we use the popular \LCDM\ cosmology whose parameters
are tabulated in Section~\ref{sec:cosmo}.  The generation of a merger
tree of collapsed halos is described in Section~\ref{sec:bmodel} and
the criteria whereby we determine which of these form star clusters is
outlined in Section~\ref{sec:model}.  

\subsection{Cosmology}
\label{sec:cosmo}

HSTC02 investigated the cooling of halos in three different CDM
cosmologies.  In this paper, we restrict our attention to the
currently-favoured \LCDM\ cosmology whose parameters are listed in
Table~\ref{tab:cosmology}.
\begin{table}
\caption{Cosmological parameters: density parameter; cosmological
constant in units of $\lambda_0=\Lambda/3H_0^2$; current baryon
density in units of the critical density; Hubble parameter in units of
$h=H_0/100$km\,s$^{-1}$Mpc$^{-1}$; power spectrum shape parameter;
root-mean-square dispersion of the density within spheres of radius
8\,$h^{-1}$Mpc.}
\label{tab:cosmology}
\begin{center}
\begin{tabular}{ccccccc}
$\Omega_0$& $\lambda_0$& $\Omega_{b0}$& $h$& $\Gamma$& $\sigma_8$\\
\hline
0.35& 0.65& 0.038& 0.7& 0.21& 0.90\\
\end{tabular}
\end{center}
\end{table}

We have used the transfer function calculated by CMBFAST.  There have
been recent suggestions that the normalisation of the power spectrum
may be closer to 0.7 than 0.9 (e.g.~\citealt{Sel02}; \citealt{ASF02})
this would have the effect of moving the formation epoch of the first
star clusters to lower redshift and also lowering the amount of
substructure.

\subsection{Block Model}
\label{sec:bmodel}

We generate a halo merger tree using the Block Model of \citet{CoK88}.
This starts with a `root' block of mass $M_0$ and density fluctuation
$\delta_0$.  In this paper we fix $M_0=10^{11}$\,\Msun\, and choose
two different values of $\delta_0$ corresponding to a 3-$\sigma$
fluctuation ($\delta_0=10.98$) and the mean density ($\delta_0=0$).

Geometrically the block can be visualised as a cuboid with sides in
the ratio 1:$2^{1/3}$:$2^{2/3}$, but the density fluctuations are
calculated as for a spherical top-hat model of the same mass; given
the uncertainty in the normalisation of the power spectrum, the
distinction is of no importance.  The block can be bisected by a plane
perpendicular to its longest axis, creating two similar blocks of half
the mass, $M_1=M_0/2$.  To generate density fluctuations in these
daughter blocks, we add power drawn at random from a gaussian
distribution with variance $\sigma^2(M_1)-\sigma^2(M_0)$; once again,
this is an approximation, but a good one.  A positive fluctuation is
added to one block and an equal negative fluctuation to the other, so
as to conserve the overall level of fluctuations in the root
block.

The same procedure is then repeated, with each parent block being
divided into two equal-mass daughters until the desired resolution is
reached.  In this paper, we use 21 levels, creating a total of
$2^{21}-1\approx 2.1\times 10^6$ blocks with a minimum block mass of
$9.5\times10^4$\Msun.

We use a simple model in which the collapse of blocks to form bound
objects is determined only by their overdensity.  To be precise, we
assume them to virialise once their linear overdensity reaches
$\delta_c=1.69$ (see \citealt{ECF96} and \citealt{LoH01} where only a
very weak dependence on $\Omega$ has been found).

The equation for the rate of growth of $\delta$ with redshift, $z$, is
\begin{equation}
\delta(z)={\delta(0)\over1+z}\,{g(\Omega)\over g(\Omega_0)},
\label{eq:delta}
\end{equation}
where
\begin{equation}
g(\Omega)=2.5\,{\Omega\over\left({1\over70} + {209\over140}\Omega -
{1\over140}\
\Omega^2 + \Omega^{4\over7}\right)}
\label{eq:g}
\end{equation}
is the growth suppression factor \citep{ViL96} and
\begin{equation}
\Omega={\Omega_0\over\left({\Omega_0+\left({1\over1+z}\right)^3\,(1-\Omega_0)}\right)}.
\label{eq:omega}
\end{equation}

In the binary tree generated by the Block Model, most of the blocks
are contained in larger blocks of greater over-density.  Often this
will be the immediate parent (one of the two daughters of each parent
will have lesser density) but it could also be a block further up the
tree.  Under these circumstances, the larger block will collapse
before the smaller, and so the latter will never attain an independent
existence as a virialised structure.  We eliminate these under-dense
blocks from the tree and call the remaining blocks `halos'.

We have performed one hundred realizations of the Block Model with
different number seeds.  We use these for both values of $\delta_0$ as
the set of halos is the same in each case.  Unless stated otherwise,
the results presented below are averages over all realizations.

\subsection{Halo evolution and the formation of star clusters}
\label{sec:model}

We wish to know whether a halo can form stars before it gets
incorporated into some larger structure.  To do this, we construct an
artificial model in which the halo has no substructure and cools at
constant density.  The actual structure of halos will be highly
complex but our model gives a reasonable estimate of what is going on,
short of doing a prohibitively time-consuming simulation.

We begin with the smallest halos and work our way up the merger tree.
Each halo is treated as an isolated, isothermal sphere, as described
in Section~3.3 of HSTC02.  The mean baryon density within the virial
radius, \rvir, is taken to be equal to
\begin{equation}
\rho_{\rm vir}= \left(\Delta_c\over\Omega\right)\rho_{b0}(1+\zvir)^3
\end{equation}
where \zvir\ is the virialization redshift, $\rho_{b0}$ is the current
mean density of baryons in the Universe, and $\Delta_c$ is the mean
overdensity of the virialized halo in units of the critical density at
that time, which we take to be $\Delta_c\approx18\pi^2\Omega^{0.45}$
\citep{ENF98}.

We define the dynamical time for each halo to be
\begin{equation}
\tdyn={1\over4\sqrt{2}}\,\tvir
\label{eq:tdyn}
\end{equation}
where \tvir\ is the age of the universe at the time of virialization.

The virial temperature of the halo is
\begin{eqnarray}
\Tvir \!\!\!\!&=&\!\!\!\! {\mu m_H\over k_B}{G\Mtot\over2\rvir} \nonumber\\
\!\!\!\!&\approx&\!\!\!\! 40.8\,{\mu\over 1.225}\,(1+z_{\rm vir})
\left(\Delta_c h^2\Omega_0\over18\pi^2\Omega\right)^{1\over3}
\left(\Mtot\over10^5\Msun\right)^{2\over3}{\rm K}.
\label{eq:tvir}
\end{eqnarray}
Here \Mtot\ is the total mass (dark plus baryonic), $m_H$ is the mass
of a hydrogen atom, $k_B$ is the Boltzmann constant, G is the
gravitational constant, and $\mu$ is the mean mass of particles in
units of $m_H$.
Note that equation~(\ref{eq:tvir}) differs slightly from the equivalent
expression in HSTC02 as the latter contains a typographical error.

The initial fractional abundance of molecular Hydrogen is taken to be
$1.1\times10^{-6}$ as calculated by \citet{GaP98}, and the initial
ionization fraction is taken to be the maximum of the equilibrium
value at \Tvir\ and the residual value from the early Universe,
$1.33\times10^{-4}$.

Starting from these initial conditions, we determine the time that it
would take the gas to cool isochorically to $\Tsf=0.75\,\Tvir$, using
the minimal model presented in Section~2 of HSTC02.  This includes
molecular hydrogen cooling, collisional excitation and ionization of
hydrogen and helium, and inverse Compton cooling from cosmic microwave
background photons.

If halos are unable to cool to \Tsf\ before being swallowed by a more
massive halo, then we assume that they are heated to the new virial
temperature and that any substructure (which would be minimal anyway
because of the long cooling time) is erased.  Contrarily, halos that
can cool to \Tsf\ are assumed to be able to (instantaneously) cool
further to low temperatures and to form a star cluster.

For halos whose cooling times are much shorter than their dynamical
times, then the gas will likely never get heated to the virial
temperature in the first place and the assumption of instantaneous
star formation will be a good one.  However, for halos in which the
cooling time exceeds the dynamical time then it seems likely that our
model will underestimate the time taken to form stars.  We consider in
Section~\ref{sec:tsfdelay} the effect of adding a time-delay before
star-formation and show that it favours the second generation of
halos.

Star clusters may or may not survive subsequent halo mergers but
either way they are assumed to instantly contaminate their
surroundings with metals.  Thus the primordial star clusters are those
that contain no smaller star clusters within them.  We assume that the
metals do not propagate into halos on other branches of the tree.
Thus metals may be ejected from star clusters but are confined within
the next level of the merger hierarchy.  This can be justified by a
self-regulated model of star formation in which star-formation is
terminated once gas is expelled from the star cluster.

We also assume that there is no external radiation field, other than
that provided by the cosmic microwave background.  Primordial star
clusters will be surrounded by neutral gas and the propagation of
ionizing photons will be severely limited.  Nevertheless, these first
objects will be highly clustered and so they will at some stage begin
to interact with each other.  We hope to consider this in a future
paper.
 
\section{Halo properties}
\label{sec:properties}

In this section, we present results for the overdense, 3-$\sigma$,
root halo.  A comparison with the mean-density root halo will be
presented in Section~\ref{sec:meanod}.

\subsection{Collapsed halos}
\label{sec:halos}

\begin{figure}
\psfig{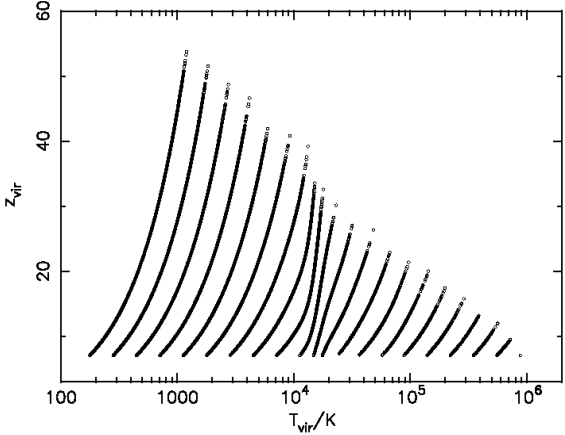}
\caption{Virialization redshift, \zvir, versus virial temperature,
\Tvir, for all halos in the 100 realisations.} 
\label{fig:zvirtvir}
\end{figure}

We start by considering the properties of all collapsed halos,
i.e.~halos whose linear overdensity exceeds that of all the
(more-massive) halos within which they are contained.  In
Figure~\ref{fig:zvirtvir} we plot the virialization redshift of such
halos (drawn from all 100 realisations) against their virial
temperature.

The banding comes about because the halos come in fixed masses.  Thus
the smallest halos of mass $9.5\times10^4\Msun$ correspond to the
left-most band; they have a wide range of virialization redshifts and
temperatures that vary between about 200 and 1200\,K; the top-most
point corresponds to a 6.2-$\sigma$ and the lowest one to a
0.9-$\sigma$ fluctuation.  20 other bands are then visible, one for
each factor of two in mass until we reach the parent halo in the
bottom-right which has a mass of $10^{11}\Msun$, a virial temperature
of $\Tvir=8.81\times10^5$K, and a virialization redshift of
$\zvir=7.1$ (corresponding to a 3-$\sigma$ fluctuation on this scale).
The bands are mostly parallel, except for the temperature range
$\Tvir\approx$10\,000--20\,000\,K within which the ionization level is
changing.

\begin{figure}
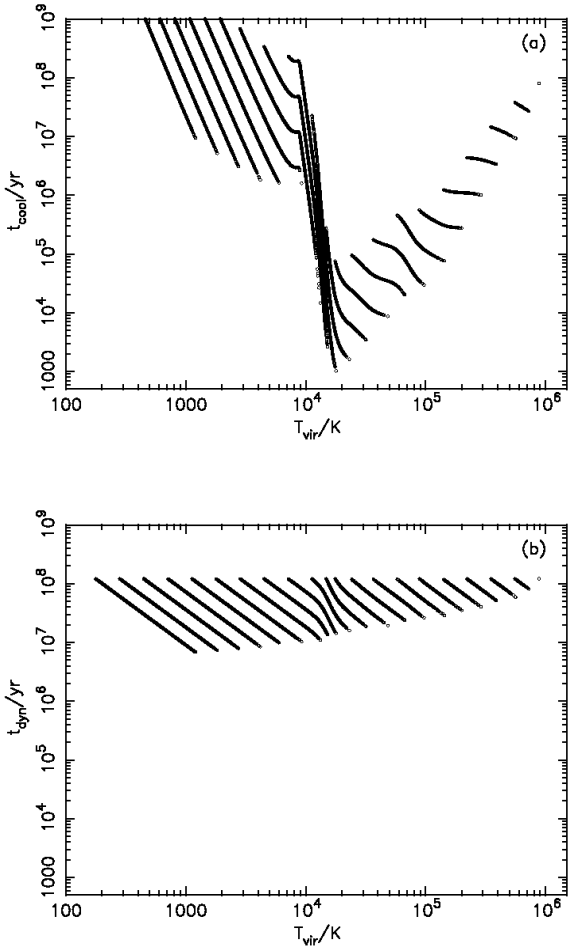

\psfig{width=8.7cm,angle=270,file=tscales_a.ps.cmp}
\psfig{width=8.7cm,angle=270,file=tscales_b.ps.cmp}
\caption{(a) The cooling time, \tcool, and (b) the dynamical time,
\tdyn, for all collapsed halos, plotted against virial temperature.}
\label{fig:talltvir}
\end{figure}

The top panel of Figure~\ref{fig:talltvir} shows the cooling time,
\tcool, versus the virial temperature for each of the halos
shown in Figure~\ref{fig:zvirtvir}.  There is a sharp decline in the
cooling time at $\Tvir\approx10\,000$\,K corresponding to the
ionization temperature of hydrogen.  Halos with higher virial
temperatures than this (\generation~2 halos) are able to cool rapidly
via electronic processes and so have relatively short cooling times.
Those with lower virial temperatures (\generation 1) have to rely on
cooling via molecular hydrogen which is formed only in very low
quantities.  Although our definition of \tcool\ only follows cooling
down to 0.75\,\Tvir, we note that \generation~2 halos have a high
residual ionization at lower temperatures that acts as a catalyst to
form molecular hydrogen: thus their cooling rates at low temperature
are faster than for \generation~1 halos.

The lower panel of Figure~\ref{fig:talltvir} shows the dynamical time
for the collapsed halos, as defined in equation~\ref{eq:tdyn}.  A a
rule of thumb, halos with virial temperatures above about 10\,000\,K
are able to cool in less than a dynamical time; cooler halos take
longer.

\subsection{Star clusters}
\label{sec:starclus}

\begin{figure}
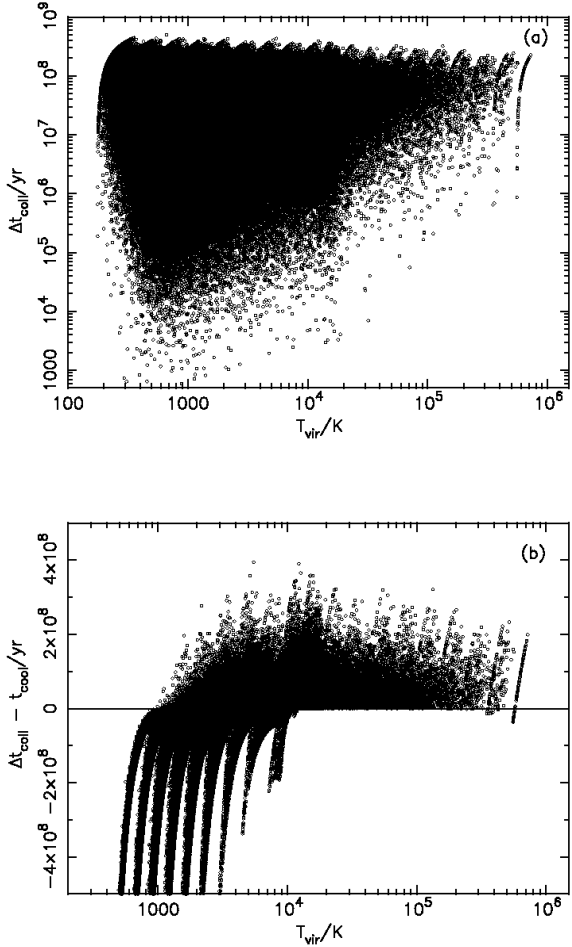

\psfig{width=8.7cm,angle=270,file=tscales_c.ps.cmp}
\psfig{width=8.7cm,angle=270,file=tpartcool.ps.cmp}
\caption{(a) The difference in collapse time of a halo and its parent,
$\dtcoll$, and (b) the difference between the collapse time of a halo
and its parent minus the cooling time of the halo, $\dtcoll-\tcool$,
versus virial temperature.  For clarity, the lower vertical axis in panel
(b) has been truncated at $-5\times10^8$yr.}
\label{fig:dtcooltvir}
\end{figure}

The time difference, \dtcoll, between the collapse time of each halo
and that of its parent is shown in the top panel of
Figure~\ref{fig:dtcooltvir}.  For a random location in space, one might
expect that the time difference would be largest for massive,
high-temperature halos.  However, that is not the case for these
realisations in which the top-level itself is constrained to collapse
at $t=0.69$\,Gyr (it is true for the halos considered in
Section~\ref{sec:meanod} for which the overdensity of the top-level
halo is zero).

Those halos for which \tcool\ is less than \dtcoll\ can cool to a
fraction 0.75 of their virial temperature before being swallowed up by
their parent halo in the merger hierarchy.  To begin with, we assume
that this is a sufficient criterion to allow them to form stars and we
identify them with star clusters.  In reality the time-delay before
star formation will be larger as the gas has to cool to low
temperatures and to congregate into regions of high-density.  We will
consider the effect of allowing a longer time-delay in
Section~\ref{sec:tsfdelay}.

The lower panel in Figure~\ref{fig:dtcooltvir} plots $\dtcoll-\tcool$
against the virial temperature.  Those halos that lie above the line
are those that form star clusters.  Just 2 percent of all collapsed
halos satisfy this condition.  However, these are not distributed
evenly over mass.  For example, in the first three levels of the
merger tree (i.e.~the three levels with the lowest mass) only a
fraction $1.2\times10^{-7}$, $8.4\times10^{-5}$ and $3.2\times10^{-3}$
of collapsed halos are able to form stars, whereas a successively
higher fraction do so at higher mass.  Only for the most massive halos
with virial temperatures in excess of $10^5\,K$ does the cooling time
again begin to exceed the lifetime of halos.

\subsection{Primordial star clusters}
\label{sec:primstarclus}

We assume that metal-enrichment from star-formation is instantaneous
but that it does not extend beyond the immediate environment of a star
cluster and its parent halo.  Then star clusters of primordial
composition are simply those which do not have any smaller star
clusters contained within them.

Approximately half of these clusters ($56.8\%$) in our 3-$\sigma$
realisations satisfy this condition with the bias swinging back
towards low masses.  The most massive primordial star cluster has a
mass of $9.76\times10^7\Msun$ and a virial temperature of
$1.54\times10^4K$.

\section{Halo mass functions}
\label{sec:massfn}

\begin{figure}
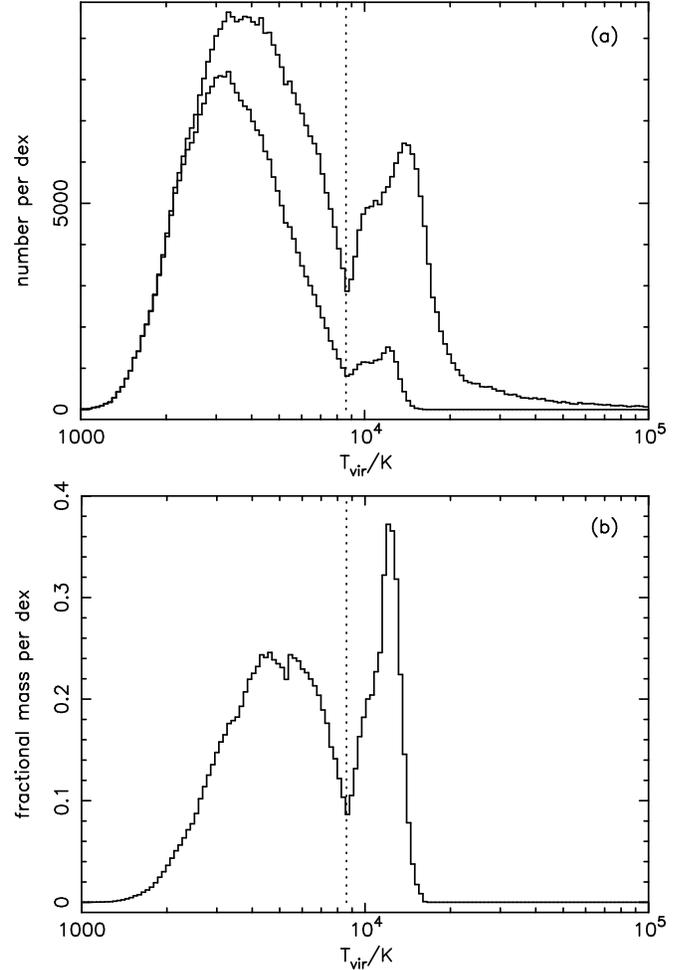

\psfig{width=8.7cm,angle=270,file=numbert.h.ps}
\psfig{width=8.7cm,angle=270,file=tempf.h.ps}
\caption{Histograms showing (a) the number of, and (b) the fractional mass
contained in, star clusters as a function of their virial
temperature.  In panel (a) the upper line shows all star clusters
whereas the low line is for clusters of primordial composition.  Panel
(b) shows only primordial halos.  The minimum at
\Tvir$\approx$8\,600\,K is used to demarcate between the two
generations of halos.}
\label{fig:histtvir}
\end{figure}

The number of star clusters as a function of virial temperature,
averaged over all 100 realisations, is shown in the upper panel of
Figure~\ref{fig:histtvir}.  The upper histogram shows all star
clusters, whereas the lower is restricted to primordial star clusters.
There is a clear minimum at about 8\,600\,K corresponding to the
division between \generation~1 halos on the left and \generation~2
halos on the right.  Note that the star clusters that make up the
upper histogram are not all independent.  That is to say that many of
the low-temperature clusters are subcomponents of the
higher-temperature ones.  However, the primordial star clusters are
all distinct objects.  The y-scale in panel (a) could be multiplied by
3.94$h^3$Mpc$^{-3}$ to convert to a number density but we have not
done this as the 3-$\sigma$ region that we consider is not
representative of all space.

The lower panel of Figure~\ref{fig:histtvir} shows the same
distribution but weighted by mass.  From this it is apparent that,
whereas the majority of the primordial star clusters in
this region are \generation~1 objects, the division of mass between
the two generations is much more even, with only about twice as much
mass contained in \generation~1 as compared to \generation~2 objects.

Note the sharp cut-off in the mass density of primordial star clusters
at virial temperatures greater than about 15\,000\,K.  This is because
all higher temperature clusters contain a \generation~2 subcluster for
which the cooling time is very short and which can itself form stars
on a short timescale.  This situation changes when we introduce a
time-delay for star formation in Section~\ref{sec:tsfdelay}.

\begin{figure}
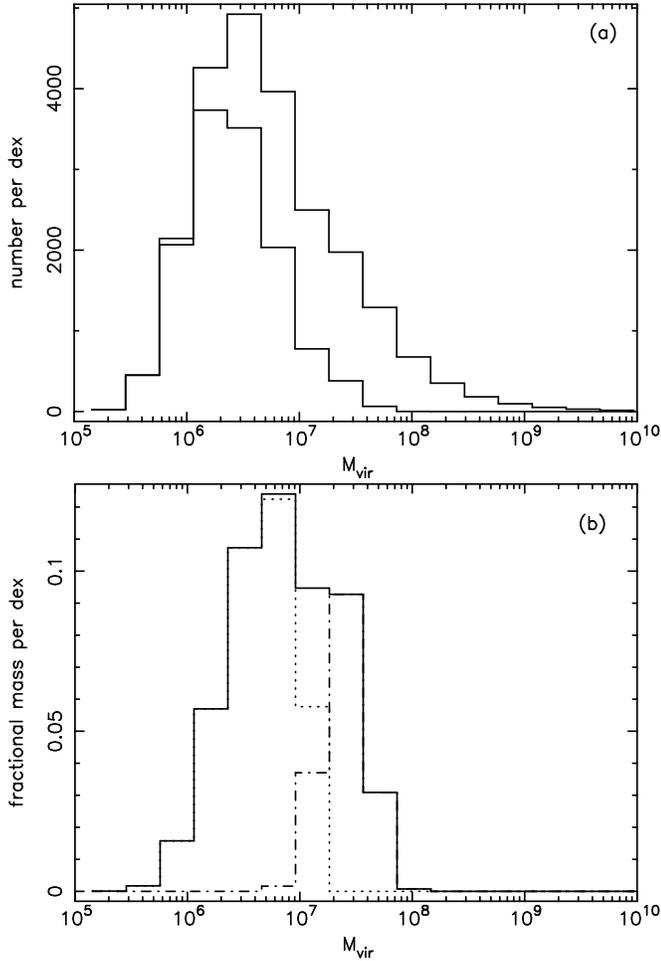

\psfig{width=8.7cm,angle=270,file=numberm.h.ps}
\psfig{width=8.7cm,angle=270,file=massf.h.ps}
\caption{Histograms showing (a) the number, and (b) the mass fraction, of star
clusters as a function of their virial mass.  In panel (a) the upper
line shows all star clusters whereas the low line is for clusters of
primordial composition.  Panel (b) shows only primordial clusters.  The
dotted line corresponds to \generation~1 halos ($\Tvir<8\,600\,$K) and the
dotted-dashed line correspond to \generation~2 halos.}
\label{fig:histmvir}
\end{figure}

Figure~\ref{fig:histmvir} is similar to Figure~\ref{fig:histtvir}
except that the ordinate is now virial mass rather than virial
temperature (to convert to baryonic mass, the x-scale should be
multiplied by 0.12).  In panel (b), the dotted line shows the
contribution to the total mass of \generation~1 halos (whose virial
temperature is less than $8\,600\,$K), while the dot-dashed line is
for \generation~2 halos.  The fractional mass contained in the two
generations is 0.109 and 0.049, respectively.  Thus about 16 per cent
of all baryons in this 3-$\sigma$ region of space will have at one
time been part of a primordial star cluster.

\begin{figure}
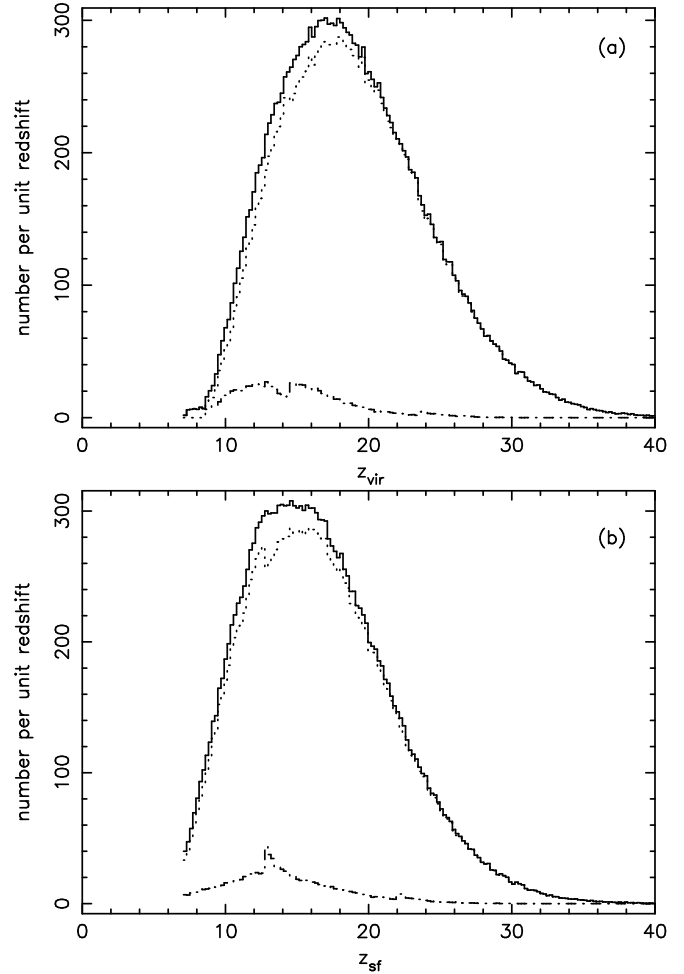

\psfig{width=8.7cm,angle=270,file=numberz.h.ps}
\psfig{width=8.7cm,angle=270,file=numberzcool.h.ps}
\caption{A histogram showing the number of primordial star
clusters as a function of (a) their virialization redshift, and (b)
their star-formation redshift.  The dash-dotted and dotted lines
correspond to \generation~1 and \generation~2 halos, respectively.}
\label{fig:histzvir}
\end{figure}

Figure~\ref{fig:histzvir} contrasts the collapse and star-formation
redshifts of both generations of halo.  The spiky features visible in
the distributions are due to the factor of two mass resolution of our
halos and would be smoothed out in a more general merger tree.
Because of the long cooling times of \generation~1 halos, the
difference in the peaks of the distributions of star-formation
redshifts of the two generations is not so great as for their collapse
redshifts.  Nevertheless, it is clear from the figure that
a significant fraction of \generation~1 halos both collapse and form
stars before \generation~2 halos begin to form in numbers.  This
highlights the need for a more sophisticated model of feedback than we
attempt in this paper.

\section{Discussion}
\label{sec:discuss}

\subsection{Numerical considerations}
\label{sec:numerics}

It is legitimate to ask to what extent our results are limited by the
factor-of-two mass resolution inherent in the Block model.  If we
could have subhalos with a wider range of masses would that lead to a
greater probability of contamination by star formation and a reduction
in the fraction of primordial halos?  The dramatic reduction in the
cooling time of \generation~2 halos compared to their low-mass,
\generation~1 subhalos, as illustrated by the upper panel in
Figure~\ref{fig:talltvir}, suggests that this is unlikely to be the
case.  We expect to move to a more realistic merger tree in future
work.  

\begin{figure}
\psfig{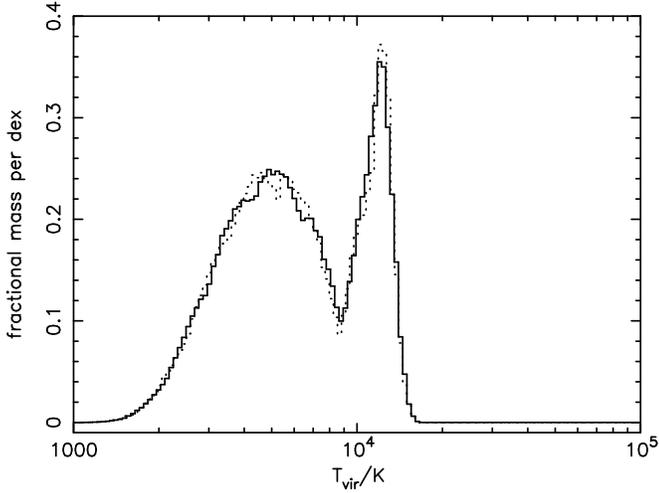}
\caption{Histogram showing the mass fraction of primordial
halos for two different choices of halo mass.  The dotted line is the
same as in Figure~\ref{fig:histtvir}\,(b) while the solid line shows
the mass fraction for a merger tree in which the mass of the root
halo has been increased a factor of $\sqrt2$.}
\label{fig:sqrt2tempf}
\end{figure}

Meanwhile, we have tested the sensitivity of our results to the
precise choice of halos masses by performing a second hundred
realizations of the merger tree with the root halo mass (and hence
each level of the merger hierarchy) increased by a factor of $\sqrt2$,
to $1.4\times10^{11}$\,\Msun.  Figure~\ref{fig:sqrt2tempf} shows the
mass function of primordial halos for these simulations contrasted
with our original simulation (dotted line).  There is no significant
difference between the two.

\begin{figure}
\psfig{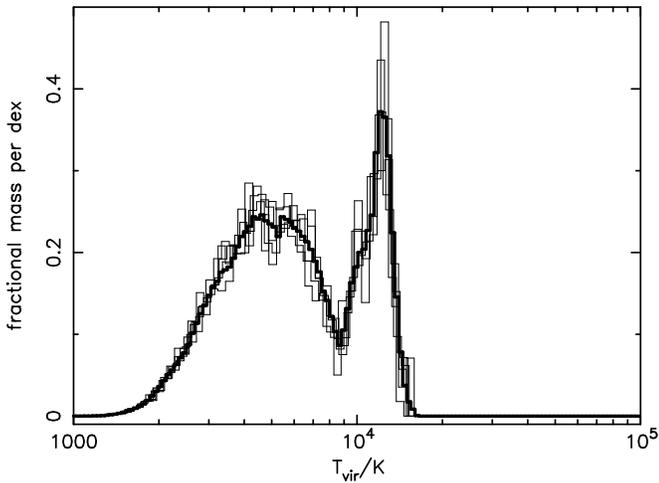}
\caption{Histogram showing the effect of different realizations on the
mass fraction of primordial objects as a function of virial
temperature.  The thick line corresponds to an average over 100
realizations.}
\label{fig:5realiz}
\end{figure}

The results that we have presented so far are an average over a large
number of realisations.  Figure~\ref{fig:5realiz} shows the dispersion
around the average, of five of the 100 realizations carried out in this
paper.   It can be seen that the scatter is significant but not enough
to seriously affect the divide between the two generations of star
clusters within each realisation.

\subsection{Time-delayed star-formation}
\label{sec:tsfdelay}

\begin{figure}
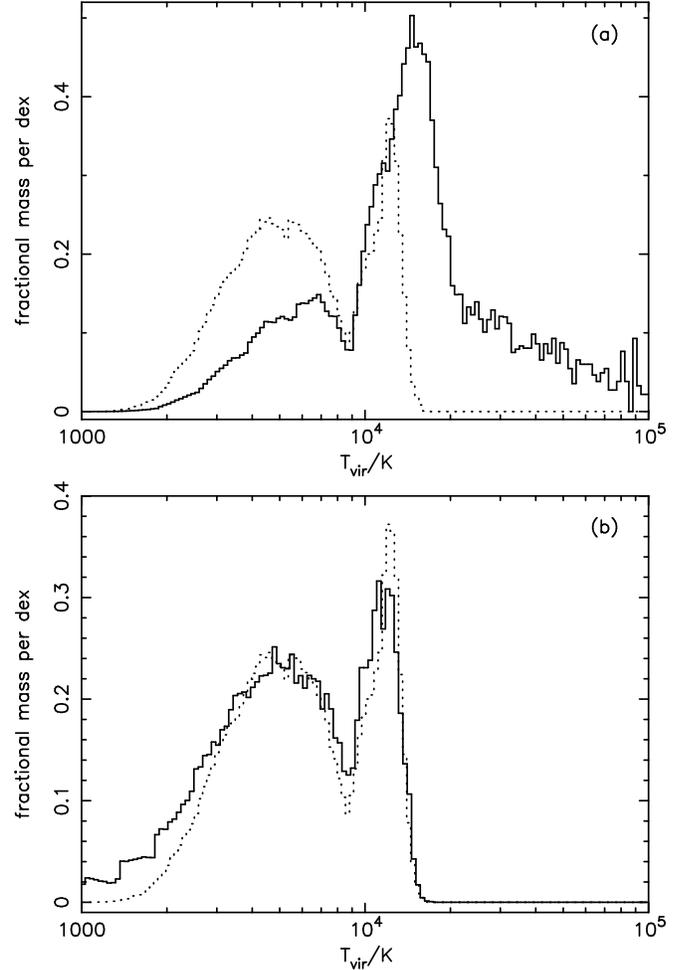

\psfig{width=8.7cm,angle=270,file=tempf.h.dynt.ps}
\psfig{width=8.7cm,angle=270,file=tempf.h.l.ps}
\caption{Histograms showing the mass fraction of primordial star clusters
as a function of virial temperature.  In Panel (a) the solid line
shows the mass fraction of halos for which a dynamical time has been
added to their cooling time.  In Panel (b) the solid line shows the
mass fraction for a zero-overdensity top-level merger tree.  In both
panels the dotted line is the same as in
Figure~\ref{fig:histtvir}\,(b).}
\label{fig:1dynt_0over}
\end{figure}

So far we have assumed that after the cooling of the gas to low
temperatures (following the T97 criterion) stars form instantaneously.
In reality, there will be a lapse of time until the gas reaches the
high-density regime in which nuclear reactions can take place and the
stars are born.  In an attempt to include in our code a time-delay
between initial cooling and star-formation, we consider in this
section the effect of adding the dynamical time to the cooling time.
The justification for this is simply that, following virialization,
one would expect the gas to take at least a dynamical time to contract
within the potential well of the halo (this argument has less force
for high-mass halos whose cooling time is very much shorter than their
dynamical time and which may therefore never attain virial equilibrium
in the first place).  Some evidence for this delayed star formation
comes from \citet{ABN02} who describe the formation of a primordial
star using 3-D hydrodynamical simulation.  Their results show that a
cooled ($\sim200\,K$) high redshift molecular cloud is formed at
$z=24$ and then a proto-star is formed at $z=18.2$.  The time that
took to form this proto-star is of the order of the dynamical time of
the cloud.

Panel (a) of Figure~\ref{fig:1dynt_0over} shows a histogram of the
fractional mass of primordial star clusters as a function of their
virial temperatures.  The dotted line corresponds to the original case
in which no time delay has been added, while the full line shows the
distribution when a dynamical time has been added to their cooling time.
The fractional mass has changed in such a way that now we have 3 times
more mass in \generation~2 halos than in \generation~1 halos,
and a greater total mass fraction than before.   Note also that there
is no longer a sharp cut-off at virial temperatures above 15\,000\,K
because it is possible for subhalos to have short cooling times in
this model and yet not to form stars.

\begin{figure}
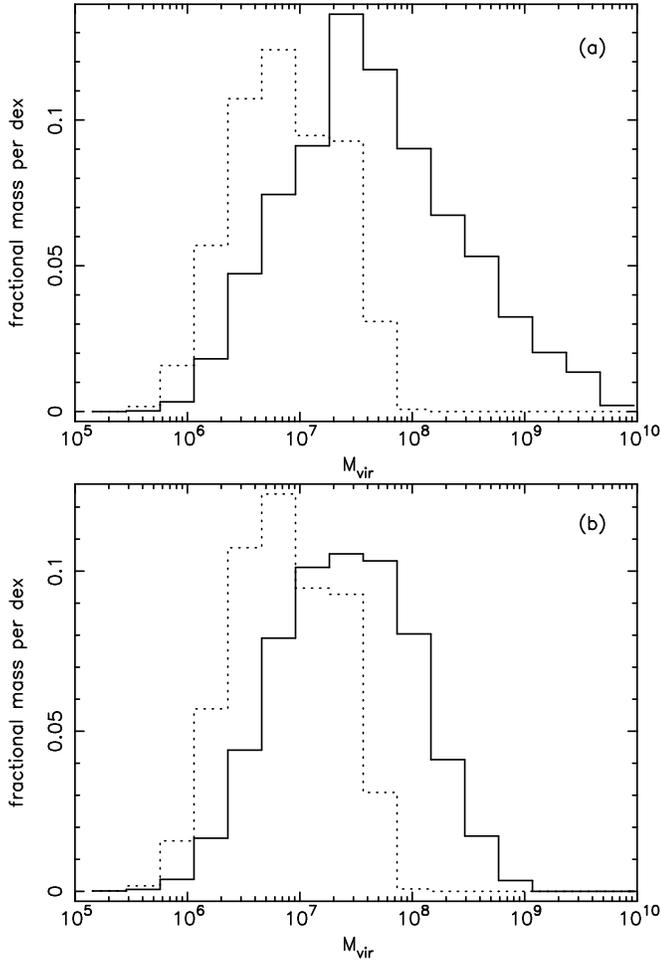

\psfig{width=8.7cm,angle=270,file=massf.h.dynt.ps}
\psfig{width=8.7cm,angle=270,file=massf.h.l.ps}
\caption{Histograms showing the mass fraction of primordial
halos as a function of their virial mass.  The dotted line in both
panels is the same as in Figure~\ref{fig:histmvir}\,(b).  In Panel (a)
the solid line shows the mass fraction of halos for which a dynamical
time have been added to their cooling time.  In Panel (b) the solid
line shows the mass fraction for a zero-overdensity top-level merger
tree.}
\label{fig:0overmassf}
\end{figure}

Panel (a) in Figure~\ref{fig:0overmassf} shows the equivalent mass
function of these clusters from which it can be seen that halos as
massive as 10$^{10}$\Msun\ can contain primordial star clusters.
While this does not seen very likely, the general conclusion that we
draw is that delayed, rather than instantaneous, star-formation will
favour \generation~2 halos over \generation~1.

We also tried a model with instantaneous star-formation but with a
time-delay before energy and metallicity feedback.  The idea is that
if the time difference between the collapse of a parent halo and the
cooling of its child is less than the time for the formation of
supernovae, $\sim10^7\,$yr, then the parent halo will be of primordial
composition and has to be added into the set of primordial objects.
However, this makes only a minor difference to our results and we will
not discuss it further.

\subsection{Mean-density regions}
\label{sec:meanod}

The first star clusters will form in overdense regions of the
Universe, hence our use of a 3-$\sigma$ root halo to this point.
However, it is interesting to contrast these results with those
expected for a more typical part of the Universe, with density equal
to the cosmic mean.  

\begin{figure}
\psfig{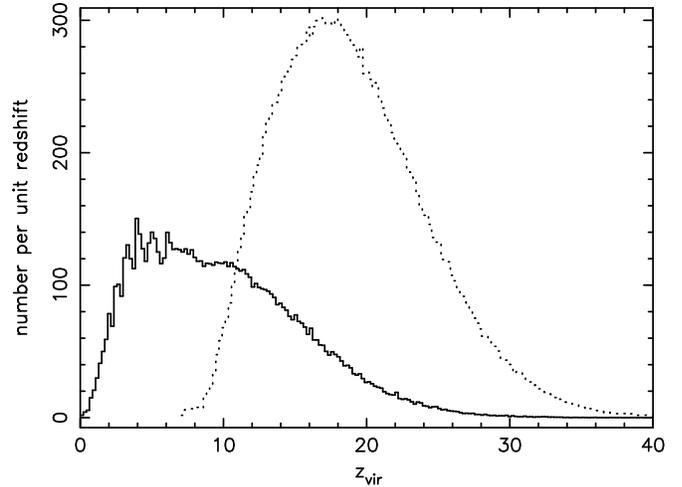}
\caption{Histogram showing the  distribution of collapse redshifts for
primordial star clusters.  The solid line is for a mean-density and
the dotted line for the previously-investigated 3-$\sigma$ regions.}
\label{fig:0overzvir}
\end{figure}

Figure~\ref{fig:0overzvir} contrasts the collapse redshifts of
primordial star clusters for the 3-$\sigma$ and mean density
regions.
It can be seen that the halos collapse at much lower redshifts in the
mean overdensity case.  Because the mean-density regions are likely to
be far-removed from the regions of the first star-formation, they are
unlikely to be affected by photoionizing photons at high redshift.
However observations \citep[e.g.~][]{FNS02} and simulations
\citep[e.g.~][]{RNA02} both suggest that the Universe became re-ionized
at a redshift of about 6 and our model will be invalid after this
time.  This will mostly affect the evolution of \generation~2 halos.

The effect of the lower formation redshift on the virial temperatures
and masses of the star clusters is shown in the lower panels of
Figures~\ref{fig:1dynt_0over} and \ref{fig:0overmassf}.  The
fractional mass distribution over virial temperature is almost
unchanged, but with a slight bias towards lower temperatures compared
with the 3-$\sigma$ case.  A greater effect is a shift in the mass
function towards higher masses: because the halos collapse at lower
redshift and hence have lower densities, they have higher masses for a
given virial temperature.

\subsection{What and where are they now?}
\label{sec:where}

Our model predicts the masses, virial temperatures and formation
redshifts of primordial star clusters, but says nothing
about their internal structure.  Hydrodynamical simulations (see
references in the Introduction) have made a start in this direction
but are as yet still in their infancy.  There has been some
theoretical speculation about the masses of the first stars but no
consensus has emerged.  In this section, we use our results to discuss
two possible fates of primordial star clusters, but note that the
physics is sufficiently uncertain that we may even have got the roles
of the two generations mixed up.

The baryonic mass of our root halos, $1.2\times10^{10}$\Msun, is
similar to that of a normal galaxy of mass approximately one-tenth that
of an $L_*$ galaxy.  The space-density for the 3-sigma
fluctuations on this scale is $3.0\times10^3h^3$Mpc$^{-3}$, similar to
that of groups of galaxies, so that we would perhaps expect one such
galaxy in a typical group.  The other galaxies will form slightly
later and so the star clusters will be biased to higher masses,
although the total mass contained in primordial star
clusters will be similar (see Section~\ref{sec:meanod}).
Our model therefore suggests that approximately one tenth of the
baryons in a typical galaxy will have passed through a primordial star
cluster.  The majority of these are probably enriched with processed
material without themselves forming a zero-metallicity star.

The majority of Population~{\sc iii} stars will be born in regions
that are destined to end up in normal galaxies.  However, our model
does not preclude the formation of some zero-metallicity stars in
low-density regions of the Universe at relatively low redshift (but
before re-ionization).  The resulting star clusters would be of
low-density and therefore very hard to detect.

The bulk of \generation~1 star clusters at high redshift have masses
in the range $10^6$--$10^7$\Msun; the baryonic mass is lower,
$10^5$--$10^6$\Msun.  It is natural, therefore, to identify these
objects with the low-metallicity globular clusters found in the bulges
and halos of normal galaxies.  The relatively long cooling times of
\generation~1 halos compared to their dynamical times would have aided
dissipative collapse within the dark matter halo and survivability of
the star cluster.  One objection to this is that zero-metallicity
stars have not been discovered in globulars, but of course the first
stars may have been of high mass and may have burnt out long ago.  A
more serious objection is that we know that the amount of material in
globular clusters is much less than one tenth of all the baryons in a
galaxy.  It is therefore probable that feedback of energy from the
first supernovae (or hypernovae) will disrupt the star clusters and
that the majority of zero-metallicity stars, should they still exist,
will be spread throughout the bulges of normal galaxies.

The main differences with \generation~2 clusters is that they are more
massive by about a factor of ten and that their cooling times are much
shorter.  It is interesting to speculate that electronic cooling when
it did turn on would lead to catastrophic accumulation of cold gas at
the centre of the collapsing halo and perhaps to the formation of a
massive black hole.  Observations (e.g.~\citealt{FeM00};
\citealt{Geb00}) give a galactic black-hole to bulge mass ratio of about
0.001.  To be consistent with this would require an accretion
efficiency of just 2 per cent, creating seed holes of mass
$2\times10^4$--$2\times10^5$\Msun.  Subsequent merging of these seed
holes could lead to the formation of supermassive black holes in the
centres of normal galaxies today.



\section*{Acknowledgements}
FS thanks his  parents for  their support; PAT is a PPARC Lecturer
Fellow.

\bibliographystyle{mn2e} 
\bibliography{early}

\end{document}